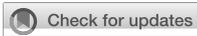

# Evaluating vision-capable chatbots in interpreting kinematics graphs: a comparative study of free and subscription-based models

Giulia Polverini* and Bor Gregorcic

Department of Physics and Astronomy, Uppsala University, Uppsala, Sweden

This study investigates the performance of eight large multimodal model (LMM)-based chatbots on the Test of Understanding Graphs in Kinematics (TUG-K), a research-based concept inventory. Graphs are a widely used representation in STEM and medical fields, making them a relevant topic for exploring LMM-based chatbots' visual interpretation abilities. We evaluated both freely available chatbots (Gemini 1.0 Pro, Claude 3 Sonnet, Microsoft Copilot, and ChatGPT-4o) and subscription-based ones (Gemini 1.0 Ultra, Gemini 1.5 Pro API, Claude 3 Opus, and ChatGPT-4). We found that OpenAI's chatbots outperform all the others, with ChatGPT-4o showing the overall best performance. Contrary to expectations, we found no notable differences in the overall performance between freely available and subscription-based versions of Gemini and Claude 3 chatbots, with the exception of Gemini 1.5 Pro, available via API. In addition, we found that tasks relying more heavily on linguistic input were generally easier for chatbots than those requiring visual interpretation. The study provides a basis for considerations of LMM-based chatbot applications in STEM and medical education, and suggests directions for future research.

KEYWORDS
generative AI, large language models, large multimodal models, chatbots, vision, STEM education, medical education, kinematics graphs

## 1 Introduction

Generative Artificial Intelligence (GenAI) is a rapidly advancing field within Artificial Intelligence (AI), distinguished by its ability to create new content, such as text, images, audio, and video. One application of GenAI is the development of chatbots, which are typically designed to engage in human-like conversations through web-based interfaces. Chatbots interact with users by interpreting input and generating contextually relevant responses. Contemporary chatbots are driven by Large Language Models (LLMs), a subset of GenAI specifically designed to generate and manipulate human language, made possible by their training on massive datasets composed of internet text, books, scholarly articles, and other written sources. Building on LLMs, chatbots are now further enhanced by Large Multimodal Models (LMMs), which extend LLMs' abilities by incorporating multiple types of data in addition to text—such as images, audio, and video—into a single model (Karwa, 2023). By leveraging both LLMs and LMMs, chatbots become highly sophisticated tools capable of interpreting and generating rich, multimodal content, thereby expanding the potential for human-AI interaction. The advent of LLM- and LMM-based chatbots has garnered considerable attention in the domain of education, sparking both excitement and concern (Ghimire et al., 2024; Onesi-Ozigagun et al., 2024) due to the technology's impressive and rapidly improving performance across a range of topics (OpenAI et al., 2024).

This paper specifically examines the ability of commercially available LMM-based chatbots to interpret user-uploaded images, also referred to as "vision." This ability is particularly relevant for education in Science, Technology, Engineering, Mathematics (STEM), and medical fields, where both experts and learners make extensive use of a wide range of visual representations, such as graphs, sketches, diagrams, etc. (Glazer, 2011). Understanding the vision abilities of these chatbots and getting a sense of what we can expect from them is becoming important for making informed decisions on their educational and professional use. Notably, several major players on the GenAI scene have publicly advertised their chatbots' vision abilities, with some even showcasing them as physics and mathematics tutors (OpenAI, 2024a; Google, 2023). Thus, we need to be informed for making decisions about the potential integration of LMM-based technology into different parts of the education process, including assessment, and curriculum design.

While several studies have explored chatbots' ability to interpret and produce language in STEM (Yik and Dood, 2024; Chang et al., 2023; Wardat et al., 2023; Vasconcelos and Dos Santos, 2023; Polverini and Gregorcic, 2024a) and medical contexts (Ghorashi et al., 2023; Sallam and Al-Salahat, 2023), research on their multimodal abilities is still less common (Polverini and Gregorcic, 2024b; Wang et al., 2024). Our study aims to contribute to the understanding of the potential of LMM-based tools in STEM and medical education fields and provide educators with the knowledge needed to make informed decisions about their potential educational uses. We approach this aim by evaluating different chatbots' performance in the interpretation of kinematics graphs, a physics topic which typically serves as a first encounter with, and a foundation for, the use of graphs in STEM fields. Our work provides an authentic assessment of the chatbots' performance, independent of commercial interests.

## 2 Background

In recent years, the use of chatbots in education has grown significantly, particularly at the high school and university levels (Gill et al., 2024; Graefen and Fazal, 2024; Dempere et al., 2023). Chatbots are being increasingly employed for a variety of educational tasks, including providing feedback (El-Adawy et al., 2024), grading student assignments (Kortemeyer, 2024), as well as tutoring (Kestin et al., 2024) and offering personalized learning experiences (Ait Baha et al., 2024). This flexibility has led to the adoption of chatbots across a wide range of educational settings (Lee, 2024; Babirye and Hanghujja, 2024; Lo and Hew, 2023).

However, the increasing uptake of chatbots in education brings with it several challenges. For example, letting students interact with third-party software involves risks related to data protection and user privacy (Sebastian, 2023). Concerns related to academic integrity (Williams, 2024; Kooli, 2023) and authentic learning (Tam et al., 2023; Parsakia, 2023) have also been highlighted, with cheating being one of the major topics under discussion. General over-reliance on chatbots can limit opportunities for deep reflection, as students may come to depend on AI-generated responses rather than developing their critical thinking and problem-solving skills (Wang et al., 2024; Wong, 2024). Further complicating this major concern is the issue of the accuracy and reliability of chatbots' generated content. While chatbot responses may appear credible, they can produce errors due to their inherent way of functioning (Polverini and Gregorcic, 2024a). In combination with a lack of students' critical interpretation, convincing but confabulated or untrue content produced by chatbots presents a real risk to student learning (Krupp et al., 2023).

In STEM and medical fields, the challenges of integrating chatbots into education are even more pronounced due to the complex and highly specialized nature of the information involved. These domains often require not just accurate answers but also the ability to engage with multiple forms of representation such as mathematical expressions, diagrams, technical drawings, and tabulated scientific data. The ability of chatbots to manage this complexity and provide reliable responses is an area that has not received much attention and requires further exploration. One of the most significant advancements in the development of chatbots is their ability to interpret non-textual input. Large multimodal models (LMMs) extend the usage of traditional text-based language models by allowing additional types of user input, such as images, audio or video. In fields like STEM and medical education, where visual representations are central to both teaching and learning, this multimodal ability represents a major step forward. However, this raises the question of how well these systems can actually interpret and generate responses from visual inputs. Education researchers have only begun to investigate this.

Our study specifically addresses this by examining the performance of LMM-based chatbots in interpreting kinematics graphs. Graphs are a foundational element for all STEM and medical disciplines, and the context of kinematics is typically the first in which students learn to deal with them systematically. As students encounter these graphs early in their learning journey, assessing how well chatbots handle visual data containing kinematics graphs is crucial for understanding their potential as educational tools. Our work aims to equip educators with the necessary knowledge to evaluate the potential role of LMM-based chatbots in their teaching practices.

### 2.1 Our previous work

In October 2023, OpenAI's ChatGPT-4 became the first publicly available chatbot with the ability to process user-uploaded images. In a previous study (Polverini and Gregorcic, 2024b), we assessed its performance on the Test of Understanding Graphs in Kinematics (TUG-K) (Beichner, 1994; Zavala et al., 2017), a well-established assessment tool designed to test students' conceptual understanding of kinematic graphs. The study revealed that, on average, ChatGPT-4 performed at a level comparable to high school students, with its performance still being far from expert-like. A qualitative analysis of the responses found that the model was prone to errors in visual interpretation, which severely impacted its overall performance on the test. On the other hand, its ability to "reason" about kinematics graphs through language was better, but somewhat unreliable and thus also not yet expert-like.

In early 2024, competitors of OpenAI launched their own vision-capable chatbots, many of which were, in contrast to ChatGPT-4, made freely accessible to users. This development extended the accessibility of this technology, particularly for users who could not afford subscriptions to "premium" versions of chatbots. In more recent work (Polverini and Gregorcic, 2024c), we expanded our study with the TUG-K to include three freely available vision-capable chatbots, Copilot (from Microsoft), Gemini 1.0 Pro (from Google), and Claude 3 Sonnet (from Anthropic), and compared their performance to the April 2024 version of ChatGPT-4. We found that, despite ChatGPT-4's performance remaining far from expert-like, it outperforms the freely available chatbots by a large margin. While the variance of performance was large across the different items on the test for all the chatbots, only one freely available chatbot's (Microsoft's Copilot) average performance was better than guessing, i.e., higher than approximately 20%.

Furthermore, in May 2024, OpenAI also released a freely accessible vision-capable version of its chatbot, called ChatGPT-4o (OpenAI, 2024b), which has not yet been tested for its performance on the TUG-K.

## 2.2 The current study

In this study, we extend our previous research by comparing the performance of several freely available and subscription-based chatbots on the TUG-K. Our findings aim to inform educational developers and teachers about the utility of various models for their educational use on tasks that involve the interpretation of graphs. Our first research question is:

*RQ1: Is the performance of subscription-based chatbots better than that of freely available chatbots?*

Our investigation of this research question aims to probe whether a performance gap exists between freely available and subscription-based chatbots, potentially indicating a new technological divide. Such a divide could have significant implications, as those who can afford subscription services might have access to superior educational support tools, which could potentially reinforce inequities in the educational space. We tested four additional chatbots (three premium and one freely available) and compared their performance against our previously collected data (Polverini and Gregorcic, 2024b; Polverini and Gregorcic, 2024c).

Our second research question relates to patterns in the different chatbots' performance on the TUG-K:

*RQ2: Can we identify any category of tasks on which the tested chatbots' performance is especially good or bad?*

We approach this question by examining if categorizing survey items based on two different categorization schemes provides any insights into what types of tasks are difficult and easy for chatbots.

Lastly, we address the performance of OpenAI's ChatGPT by looking at its evolution from October 2023 onwards, and comparing the results to its latest version, ChatGPT-4o, released in May 2024. Our last research question is thus:

*RQ3: How has OpenAI's vision-capable chatbots' performance on the TUG-K evolved since October 2023?*

Answering this question provides an overview of the performance of the state-of-the-art and best-performing chatbot family, both in terms of its current state and how it has evolved since the first release of the vision-capable version of ChatGPT.

## 3 Method

Our methodology is consistent with that adopted in our previous research on the topic (Polverini and Gregorcic, 2024b; Polverini and Gregorcic, 2024c), to allow comparisons and facilitate discussion.

## 3.1 The test of understanding graphs in kinematics (TUG-K)

Learning science subjects requires an understanding of key concepts. Concept inventory assessments are research-based, typically multiple-choice tests, which measure understanding of selected concepts, such as evolution (Perez et al., 2013), Newtonian force (Hestenes et al., 1992), and many others. They represent valuable tools, allowing educators insights into student understanding. Moreover, they have often been found to reveal difficulties in students' post-instruction comprehension, especially in traditionally structured courses (Furrow and Hsu, 2019). The results from these surveys can inform educators and allow them to design more meaningful and effective teaching methods.

One such concept inventory assessment is the Test of Understanding Graphics in Kinematics (TUG-K), originally designed by Beichner (1994) and last updated by Zavala et al. (2017). It consists of 26 items that aim to assess students' ability to work with graphical representations of kinematics concepts. Every item consists of a question, a graph depicting the task's scenario, and five multiple-choice answer options, which could be numerical results, other graphs, strategies for approaching tasks, descriptions, etc.[1]

Students' common difficulties in interpreting kinematics graphs are well-documented (e.g., (McDermott et al., 1987; Bowen and Roth, 1998; Bragdon et al., 2019)), and research exists on students' performance on the TUG-K (Zavala et al., 2017). In our previous works and this study, we use the TUG-K as a benchmark for testing chatbots' performance in interpreting kinematics graphs. The different chatbots' performance on this test can be seen as indicative of their ability to deal with reading and interpreting graphs.

## 3.2 The selection of chatbots

In our initial work involving the TUG-K, we tested the subscription-based ChatGPT-4 in October 2023, immediately after its vision abilities were made available to end users (Polverini and Gregorcic, 2024b). Between March and April 2024, we extended the testing to freely available chatbots (Polverini and Gregorcic, 2024c). We counted a chatbot to be freely available if it allowed a limited or unlimited number of prompts to users who were not paying a subscription fee. However, they may still require registering as a user. The tested freely available chatbots were: Anthropic's Claude 3 Sonnet, Microsoft's Copilot, and Google's Gemini 1.0 Pro.

In May, we also tested the newest and most advanced OpenAI's model, ChatGPT-4o, which is expected to become freely available for all users, as per OpenAI's press release (OpenAI, 2024b). ChatGPT-4o is likely to be OpenAI's response to other players making chatbots with vision capabilities freely available earlier in spring 2024.

---

1  In order to facilitate the reading and interpretation of this paper, we suggest readers download the whole test from the Physport website (AAPT, 2017).

In April 2024, we also collected new data with ChatGPT-4, to capture the performance of the updated version of the chatbot and to allow the comparison of the different chatbots' performance at roughly the same point in time (spring 2024). The most noteworthy upgrade to the subscription-based ChatGPT-4 since October 2023 is the integration of Advanced Data Analysis mode (ADA), a plug-in that allows the chatbot to execute data interpretation and visualization by running Python code within the chat interface. To test if the use of ADA impacts ChatGPT-4's performance on the test, we collected data with and without the ADA plug-in enabled. However, when comparing different chatbots' performance, we used ChatGPT-4's data without the ADA plug-in (Polverini and Gregorcic, 2024c). For more details on how ADA impacts ChatGPT-4's performance, see section 4.3.

To examine how subscription-based models stack against the freely available ones, we tested Claude 3 Opus, the premium version of Claude 3, and Gemini 1.0 Ultra, the premium version of Gemini. Furthermore, we also tested Gemini 1.5 Pro, which is Google's latest model accessible through the Gemini API. This is the only model that we do not test through a chatbot interface. However, because it is likely to become more widely accessible in the future, we deemed it appropriate to include it in the study.

To summarize, we investigated the performance of eight LMM-based chatbots: four freely available (Copilot, Gemini 1.0 Pro, Claude 3 Sonnet, and ChatGPT-4o) and four subscription-based (ChatGPT-4, Claude 3 Opus, Gemini 1.5 Pro, and Gemini 1.0 Ultra).

## 3.3 Testing procedure

We took screenshots of each of the 26 individual test items and uploaded them as an attached "png" image to the prompts. The way in which an LMM is prompted deeply influences the output (Polverini and Gregorcic, 2024a). In order to not introduce prompt-related bias to our data and ensure that the responses were based on the uploaded image data only, we aimed to use only the screenshot itself as the prompt. However, this was not always possible. Google's Gemini 1.0 Pro, Ultra, and Microsoft's Copilot do not allow image-only prompting, so in those cases we had to include a textual prompt. To exclude the possibility of the models gleaning important information from the textual prompt, we opted to use the following minimalist instruction: "Answer the question in the image."

The prompt was repeated several times to take into account the probabilistic nature of the generation of responses by the chatbots. From our original work (Polverini and Gregorcic, 2024b), in which we prompted every item 60 times, we learned that a clear and robust distribution of responses emerges in much fewer iterations. We thus opted to repeat each item 30 times, resulting in 780 responses from each chatbot. This number allows for a balance between time-consuming data collection and an acceptable degree of uncertainty in the results. Figure 1 shows an example of a conversation window in ChatGPT, with a TUG-K item as a prompt, followed by the chatbot's response.

Each prompt was submitted in a new conversation window, rather than using the regenerate option within the same chat. This choice was motivated by the fact that chatbots base new responses on previous text in the conversation, also referred to as the context (Polverini and Gregorcic, 2024a). By submitting the prompt in a new chat every time, each repetition of the prompt is treated as a completely new task. No further parameters (such as length of the expected response, temperature, etc.) could be changed in the chatbot window or the API settings.[2] The only exception is Microsoft's Copilot, which could be set to one of the three predetermined modes: creative, balanced, or precise. We opted to use the "balanced" setting. The probabilistic behaviour of chatbots allows us to treat the data as "synthetic samples" of responses. However, it is important to note that such a sample does not reflect any particular person's or population's understanding of the topic. Rather, it reflects patterns emerging from the diverse data used for training the models.

The responses were collected by copy-pasting the screenshots of the survey items (and in the case of Gemini and Copilot the one-sentence text instruction) into the chatbot interface, submitting it to generate the response, and immediately copy-pasting the output into a local repository text file. This was done by both authors of the paper on a rolling basis during the approximately 3-month period of data collection from March to May 2024. Because some chatbots had limitations on the number of prompts in a given time window (typically reset every 3 or 4 h), the authors also prompted on weekends and late evenings, often prompting several chatbots in parallel. With some chatbots (i.e., Gemini and Copilot), which had more generous usage limits, the prompt limitation was not an issue.

## 3.4 Data analysis approach

None of the outputs from different chatbots contained only a letter as the response to the question in the image. The chatbots tended to respond by describing the task, proposing a strategy to solve it, or justifying the selection of the answer options. This indicates that the chatbots are possibly fine-tuned or given a system prompt, which facilitates them elaborating on their answers. In some cases, the chatbots followed what appears to be a Chain-of-thought approach (Wei et al., 2023), as can be seen in Figure 1. However, analysing the whole response in terms of the content of the text accompanying the final answer would require an in-depth qualitative analysis, similar to that done in our previous work (Polverini and Gregorcic, 2024b). While we intend to perform such an analysis in the future, this paper aims to present an overview on a quantitative level, focusing solely on the final answers.

The chatbots' responses were predominantly structured to facilitate their rapid coding. While the chatbots' responses often contained argumentation for the selected letter option, the chosen answer was in the vast majority of cases clearly stated at the beginning or the end of the response, often in bolded letters. Coding most responses was trivial and fast, typically requiring less than a minute per response. After the first round of coding, where both authors coded all the responses independently, the inter-coder agreement was 97.4 to 99.6%, depending on the chatbot. After discussing the coding discrepancies, a total agreement was reached. The total time spent coding all 6240 collected responses was about 10 full working days for each coder.

---

2  The Gemini 1.5 Pro model was prompted using the console in Google AI studio, using the default settings, which could not be changed at the time of data collection.

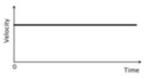

FIGURE 1
The figure shows a screenshot of a TUG-K item (top) and a conversation window in ChatGPT-4o (bottom) where the said item screenshot was uploaded as the prompt. In this case, the chatbot gave a correct response.

In cases where the chatbots did not select one of the five options clearly, we considered the questions not answered and coded the responses as "N." The proportion of N-coded responses ranged from 8 to 15%, depending on the chatbot, with most chatbots being around 10%. Alternatively, when a chatbot did not pick a letter but clearly stated in words the answer marked with the said letter, we coded the associated letter option as its choice. This was most common for Claude 3 Opus (9.0%) and Gemini 1.0 Ultra (6.3%), but otherwise rare (0–2.4%).

We have publicly shared our research data in an online data repository to ensure transparency and data availability (see Data availability statement).

The answers coded according to the selected letter option were entered into a table, compared to the answer key and marked as correct or incorrect. N-coded responses were counted as incorrect.

# 4 Findings

## 4.1 The overall performance

Table 1 presents the performance of the eight tested chatbots. The freely available chatbots include Claude 3 Sonnet, Gemini 1.0 Pro, Copilot, and ChatGPT-4o; the subscription-based models are Gemini 1.0 Ultra, Claude 3 Opus, Gemini 1.5 Pro API, and ChatGPT-4. In the first column, the TUG-K items are sorted from the best average performance across all chatbots (item 19) to the lowest (item 1). The LMMs are sorted left to right, according to their average performance on the test as a whole, with Gemini 1.0 Ultra being the lowest performing and ChatGPT-4o being the best.

Among the freely available models, Gemini 1.0 Pro and Claude 3 Sonnet exhibit particularly poor performance, with average scores below 20, and 0% scores on 14 and 12 items, respectively. Copilot performs slightly better, scoring 0% on 8 items. This is somewhat surprising, considering that both Copilot and ChatGPT-4 rely on the same underlying model, GPT-4. The difference in their performance likely lies in the adaptations made by their developers, Microsoft and OpenAI respectively, in developing the chatbot application, such as fine-tuning and system prompts. ChatGPT-4o performs much better, outperforming all other freely available chatbots and surpassing its subscription-based counterpart by 16.2 percentage points. This is a case against the expectation that subscription-based models would be superior to the freely available ones, challenging the idea that paying for access to "premium" models guarantees better performance.

The expectation that subscription-based models would outperform freely available ones is further contradicted by the fact that Gemini 1.0 Ultra performs worse than all freely available versions, including Gemini 1.0 Pro. We can also see that Claude 3 Opus is only marginally better than Claude 3 Sonnet, both performing worse than Copilot. While Gemini 1.5 Pro does better than the earlier models, it still falls short of ChatGPT-4 by almost 8 percentage points and of ChatGPT-4o by almost 24 percentage points.

It is also interesting to note that the four lowest-performing chatbots (Gemini 1.0 Ultra, Claude 3 Sonnet, Gemini 1.0 Pro, and

TABLE 1 Performance of the selected chatbots in percentage of correct responses, ranked from lowest to highest overall score.

| Item | Gemini 1.0 Ultra | Claude 3 Sonnet | Gemini 1.0 Pro | Claude 3 Opus | Copilot | Gemini 1.5 Pro API | Chat GPT 4 | Chat GPT 4o | Avg |
|---|---|---|---|---|---|---|---|---|---|
| 19 | 80 | 93.3 | 100 | 96.7 | 100 | 100 | 93.3 | 100 | **95.4** |
| 10 | 10 | 66.7 | 0 | 76.7 | 96.7 | 96.7 | 100 | 100 | **68.3** |
| 6 | 0 | 83.3 | 83.3 | 53.3 | 40 | 63.3 | 53.3 | 63.3 | **55** |
| 20 | 0 | 76.7 | 0 | 16.7 | 0 | 93.3 | 100 | 93.3 | **47.5** |
| 3 | 0 | 0 | 0 | 40 | 60 | 100 | 50 | 100 | **43.8** |
| 22 | 100 | 23.3 | 0 | 6.7 | 26.7 | 6.7 | 83.3 | 100 | **43.3** |
| 16 | 3.3 | 13.3 | 80 | 10 | 30 | 70 | 76.7 | 13.3 | **37.1** |
| 26 | 0 | 36.7 | 0 | 16.7 | 36.7 | 0 | 86.7 | 100 | **34.6** |
| 2 | 13.3 | 3.3 | 0 | 16.7 | 63.3 | 60 | 63.3 | 50 | **33.8** |
| 25 | 0 | 0 | 0 | 20 | 0 | 100 | 40 | 100 | **32.5** |
| 4 | 30 | 43.3 | 83.3 | 10 | 20 | 13.3 | 36.7 | 10 | **30.8** |
| 18 | 0 | 0 | 0 | 26.7 | 50 | 56.7 | 26.7 | 66.7 | **28.3** |
| 12 | 0 | 0 | 0 | 13.3 | 86.7 | 6.7 | 56.7 | 40 | **25.4** |
| 11 | 60 | 6.7 | 53.3 | 3.3 | 3.3 | 23.3 | 20 | 33.3 | **25.4** |
| 17 | 0 | 0 | 0 | 0 | 0 | 33.3 | 43.3 | 100 | **22.1** |
| 13 | 26.7 | 3.3 | 3.3 | 46.7 | 0 | 13.3 | 10 | 56.7 | **20** |
| 21 | 10 | 6.7 | 10 | 23.3 | 3.3 | 3.3 | 33.3 | 60 | **18.8** |
| 5 | 36.7 | 6.7 | 13.3 | 30 | 16.7 | 30 | 3.3 | 0 | **17.1** |
| 8 | 0 | 3.3 | 0 | 0 | 20 | 0 | 10 | 100 | **16.7** |
| 24 | 0 | 0 | 0 | 3.3 | 0 | 3.3 | 26.7 | 100 | **16.7** |
| 7 | 3.3 | 0 | 13.3 | 6.7 | 10 | 3.3 | 13.3 | 43.3 | **11.7** |
| 23 | 0 | 0 | 20 | 36.7 | 0 | 0 | 0 | 33.3 | **11.3** |
| 14 | 6.7 | 0 | 0 | 16.7 | 3.3 | 0 | 36.7 | 16.7 | **10** |
| 15 | 20 | 0 | 13.3 | 6.7 | 3.3 | 16.7 | 0 | 20 | **10** |
| 9 | 0 | 0 | 16.7 | 10 | 0 | 3.3 | 40 | 0 | **8.8** |
| 1 | 6.7 | 0 | 0 | 6.7 | 0 | 6.7 | 0 | 23.3 | **5.4** |
| Avg | **15.6** | **17.9** | **18.8** | **22.8** | **25.8** | **34.7** | **42.4** | **58.6** | **29.6** |

The blue cells highlight the best performance on each item; the red cells represent the lowest performance on each item.

Claude 3 Opus) perform about at the level of guessing (15.6–22.8%), while Copilot performs only marginally better (25.8%).

*Answering RQ1: Is the performance of subscription-based chatbots better than that of freely available chatbots?*

Counter to our initial expectations, we have not found that subscription-based chatbots generally outperform freely available ones. While the subscription-based ChatGPT-4 outperforms the freely available Copilot (both are based on the GPT-4 model), the freely available ChatGPT-4o outperforms both. Comparing the subscription-based and freely available versions of Gemini and Claude 3, we see very little differences, with Claude 3 Opus outperforming Claude 3 Sonnet only marginally, while the subscription-based Gemini 1.0 Ultra actually performs worse than the freely available Gemini 1.0 Pro. The Gemini 1.5 Pro model, available to subscribing users through Google's AI studio as an API, outperformed both Gemini 1.0 chatbots, but is currently not available in the form of a chatbot application.

## 4.2 Looking for patterns

There is a lot of variance in chatbot performance across different items. For example, item 19 has high scores across the board (ranging from 80 to 100%), while item 1 has universally low scores (ranging from 0 to 23.3%). This suggests that certain tasks are challenging across all models, while others are, on average, easier. However, it is not clear what makes a task difficult or easy for a chatbot.

Here we use the data from all eight chatbots to examine some possible patterns in their performance. To do this, we categorize the survey items in two ways: (1) based on content-based knowledge objectives addressed by items on the test, and (2) based on the procedures required to answer the items on the test.

### 4.2.1 Categorizing by content-based knowledge objectives

First, we apply the categorization according to areas of conceptual understanding for which the test was developed. This categorization is given by Zavala et al. (2017) and divides the items into seven content-based groups, referred to as different test objectives. A detailed presentation of the objectives is provided in Supplementary material, part A, together with a table of chatbots' performance, grouped by objectives.

For example, items 2, 6, and 7 test the ability to determine the acceleration from a velocity graph. Item 2 requires determining the interval with the most negative acceleration, while the other two ask to determine the negative (item 6) and positive (item 7) values of velocity at a given time. While the objectives are clear, well defined, and make sense from a conceptual and educational perspective, the variance of item difficulty within the test objective groupings is often large for individual chatbots, as well as across different chatbots.

Data for Objective 3 (*Determine the change of position in an interval from the velocity graph*) exemplify this clearly. The average performance of different chatbots on the objective ranges from 30% to 60.8%, while individual chatbots' performance on the items in this objective ranges from 0% to 100%.

Examining the graphs in Figure 2, we can see that the internal range (the range of performance of a given chatbot in a given objective) is more than 40 percentage points for most combinations of chatbots and objectives. There are some cases in which the spread of performances of individual chatbots is less than 20 percentage points indicating a stable performance on a given objective. However, this mostly occurs in objectives where the performance is poor (20% or less). Given that the performance of most chatbots on the test as a whole is poor, it is very likely that performance on some or several objectives will also be poor. This would also be the case if we randomly grouped items into arbitrary groups. We can also see that ChatGPT-4 and ChatGPT-4o perform noticeably better than most other chatbots, but the range of their performance on individual objectives is also mostly large, with only a handful of exceptions.

Thus, from the available data, it appears that grouping the items in terms of test objectives does not provide novel meaningful insights into the strengths and weaknesses of individual chatbots. Looking at all chatbots together, while there are differences in the average performance across objectives, we also cannot make generalizable conclusions because of the large range of the different chatbots' performance on each objective.

### 4.2.2 Categorizing by procedural nature of items

Another possible way of grouping the items is according to the procedural nature of the task itself. We identified eight different categories of tasks. Supplementary material, part B, provides a detailed presentation of the categories grouped by the procedure required to solve the task, together with a table of the results ranked by task procedure.

For example, items 5, 6, 7, and 18 (Procedure II) ask for determining a kinematics quantity (velocity, acceleration) by determining the slope of a given graph. The available answer options provide possible numerical results. This categorization focuses more on the procedural aspects of answering the item than on the underlying kinematics concept knowledge.

Figure 3 shows the performance of the different chatbots in each of the 8 procedural categories, referred to as Procedure I to VIII. Similarly to the previous categorization system, we see a large range in individual chatbots' performance within the same procedural category, as well as a large range in average performance across the different chatbots.

The two categories that seem to exhibit the most robust pattern are Procedure I (*Select a suitable strategy*) and Procedure VI (*Determine the slope of a graph at a given time*). However, even in these categories there are outliers (such as Gemini 1.0 Ultra and Gemini 1.0 Pro) that perform very low at task 10 (Procedure I), at which all other chatbots perform between 67 and 100%. Interestingly, both of these chatbots perform much better at task 19, which suggests that there are other important factors at play besides the overall procedural character of the task. Despite these differences, items 10 and 19 remain those with the best overall performance across, highlighting a result that was already emerging from our previous study: tasks that require providing a strategy for solving the problems are, in general, easier for chatbots than those looking for an actual comparison or interpretation of graphs (Polverini and Gregorcic, 2024b). This means that chatbots tend to be more successful when they are asked to deal with language rather than images (Gregorcic et al., 2024; Yeadon et al., 2023).

Procedure VI (*Compare surface areas under graphs and select the one with the largest surface area*) stands out as a relatively consistently difficult one for all the chatbots. The range of average performance on the items in this category is the narrowest of all the eight procedural categories (8.3%), with individual chatbots also exhibiting relatively small differences in performance on the items within the category.

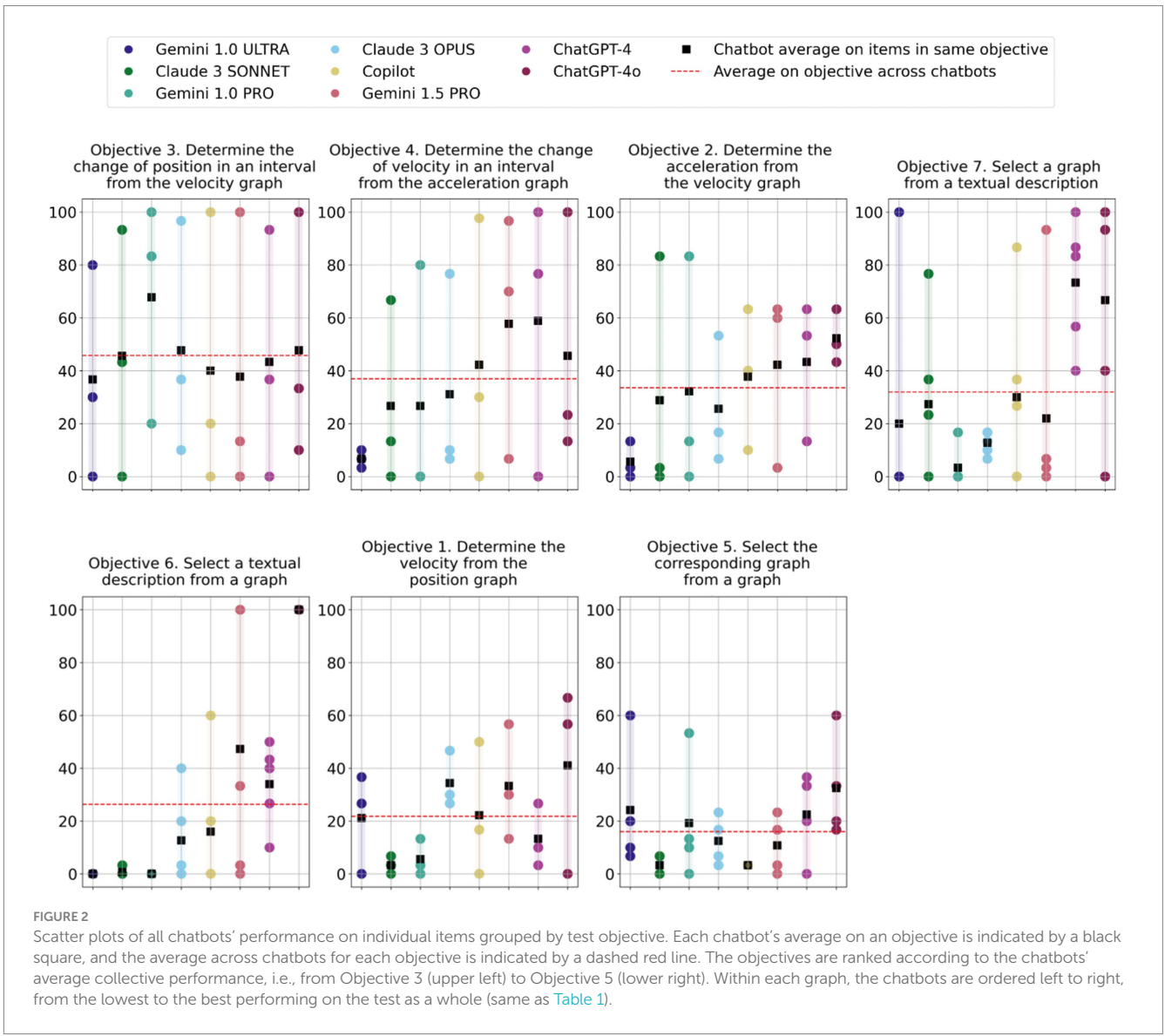

FIGURE 2
Scatter plots of all chatbots' performance on individual items grouped by test objective. Each chatbot's average on an objective is indicated by a black square, and the average across chatbots for each objective is indicated by a dashed red line. The objectives are ranked according to the chatbots' average collective performance, i.e., from Objective 3 (upper left) to Objective 5 (lower right). Within each graph, the chatbots are ordered left to right, from the lowest to the best performing on the test as a whole (same as Table 1).

*Answering RQ2: Can we identify any category of tasks on which the tested chatbots' performance is especially good or bad?*

The two different ways of categorizing the survey items provide little useful insight into generalizable principles of what makes a task easy or difficult for a given chatbot or our entire cohort of chatbots. One conclusion that we can draw is again that chatbots overall perform better on linguistic tasks than on those that require more advanced vision abilities. This is in agreement with previous research on the topic (Polverini and Gregorcic, 2024a; Polverini and Gregorcic, 2024b; Polverini and Gregorcic, 2024c). The second conclusion we can draw from our data is that comparing the sizes of surface areas under different graphs presents a challenge to most chatbots, although it is unclear if there is one single reason for this difficulty. A more detailed qualitative analysis of the chatbots' responses could potentially provide more insights.

## 4.3 Comparisons between different versions of ChatGPT

ChatGPT-4 and ChatGPT-4o even more so, outperform other chatbots in the TUG-K. However, it is interesting to examine the performance of different versions of ChatGPT. Here, we compare four versions: the first released version of ChatGPT-4 with vision capabilities from October 2023, its April 2024 version, its April 2024 version with enabled use of the Advanced Data Analysis (ADA) plug-in, as well as the latest version powered by the new GPT-4o model, released in May 2024. Table 2, compares the different versions of OpenAI's chatbot.

A key finding is that overall performance between the October and April versions is similar. However, most items show changes

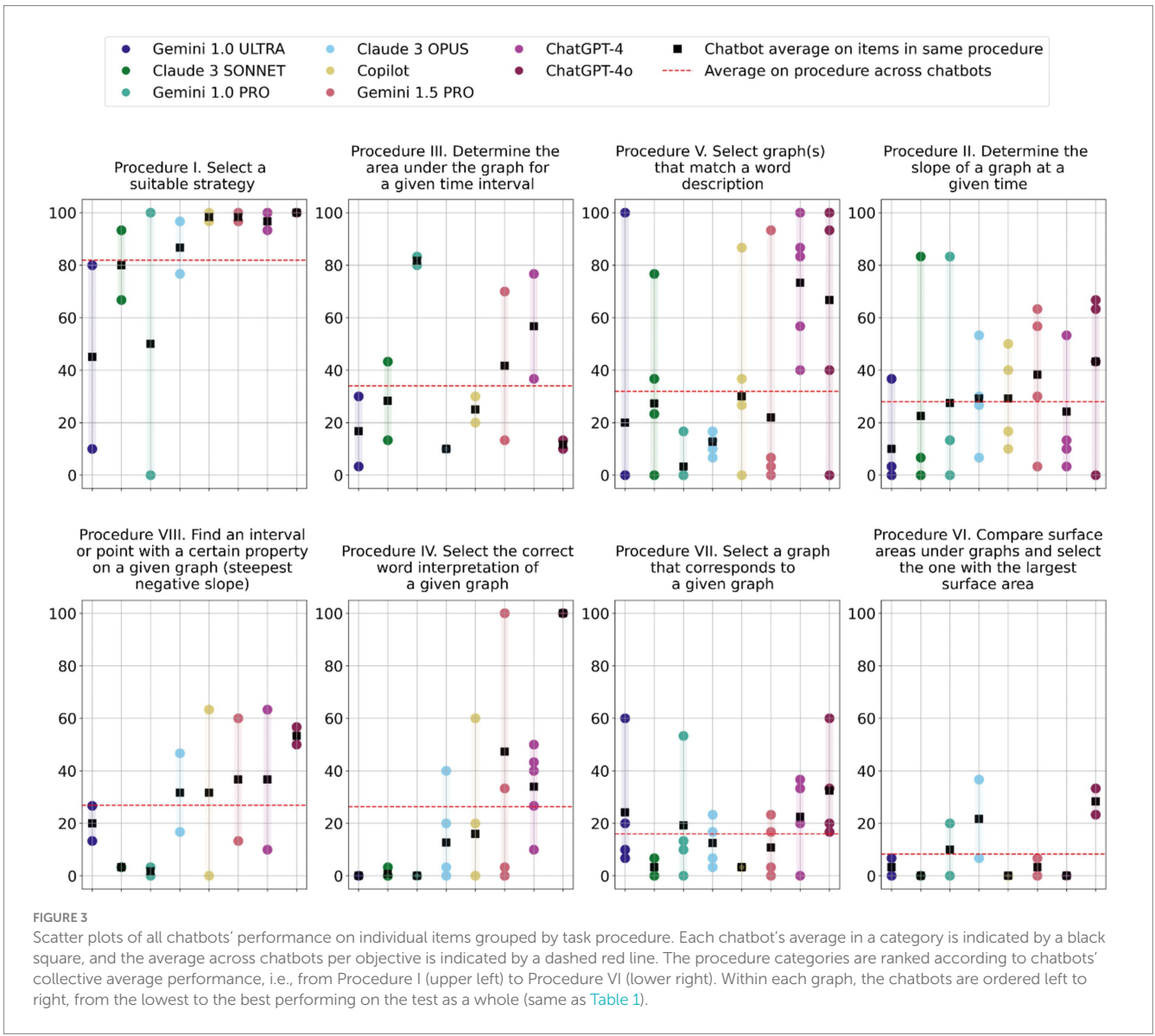

FIGURE 3
Scatter plots of all chatbots' performance on individual items grouped by task procedure. Each chatbot's average in a category is indicated by a black square, and the average across chatbots per objective is indicated by a dashed red line. The procedure categories are ranked according to chatbots' collective average performance, i.e., from Procedure I (upper left) to Procedure VI (lower right). Within each graph, the chatbots are ordered left to right, from the lowest to the best performing on the test as a whole (same as Table 1).

in performance, sometimes improvements, sometimes deteriorations, without a clear pattern or identifiable reason. This unpredictability highlights the variable nature of the model's performance and sensitivity to model updates and fine-tuning (Chen et al., 2024; Luo et al., 2024). Notably, we saw a qualitative difference in the responses coded as "N" (not answered). In both the October and April versions, 8% of responses were coded as "N." While in October the code was mainly related to statements in which the chatbot claimed it could not solve the task because of a lack of information[3] (Polverini and Gregorcic, 2024b), in April the "N" code was also due to the chatbot asking for confirmation or further instructions and thus not giving an answer in its response. For instance:

---

3 The other reason for coding the output as "N" was the ambiguity of the responses, since sometimes the chatbot could not decide which option was the correct one among two possibilities.

"[…] If you have more precise data points from the graph or if you can zoom in on the graph to read the values more accurately, I could assist you in calculating a more accurate slope. Otherwise, it's challenging to provide an exact answer from the given image alone. Would you like to proceed with a different approach?" [ChatGPT-4 April 2024 No ADA, response 6–18].

"[…] Comparing this description with the provided velocity vs. time graphs, we can determine which one correctly matches the acceleration graph. Would you like me to describe which graph (A-E) matches these conditions?" [ChatGPT-4 April 2024 No ADA, response 15–08].

It appears that the April 2024 version of the chatbot is less likely to directly provide an answer and more often turns to the user in a manner characteristic of a collaborating partner. Gregorcic et al. (2024) have previously suggested that ChatGPT has evolved to become a more cooperative partner for engaging in a productive dialogue.

Another interesting aspect is the integration of the ADA plug-in into the ChatGPT-4 interface, enabling the chatbot to connect to a Python code interpreter. This plug-in was utilized automatically in the 70% of responses belonging to 6 out of the 26 items, for tasks where the chatbot decided its use to be warranted. Contrary to our expectations, the plug-in did not consistently improve performance on the TUG-K. In fact, the ADA led to slight declines in performance on three items, the same performance on two, and only slight improvement on one. A closer look at the responses involving the ADA shows that the plug-in was used ineffectively and was often fed with incorrect information on which it then performed calculations. 66.7% of responses which used ADA were coded "N." In addition to the above-mentioned reasons, an important contribution to the high proportion of "N"-coded responses was that when using the ADA the chatbot often asked the user to confirm further use of it, instead of directly answering the question, for example:

> "If you would like a more accurate determination, I can perform a graphical analysis to estimate the slope. Would you like me to do that?" [ChatGPT-4 April 2024 ADA, response 7–05].

Table 2 also allows us to compare ChatGPT-4o's performance to the different versions of ChatGPT-4. It outperforms the April 2024 version of ChatGPT-4 (without ADA) by 16.2 percentage points. Performance has improved on 17 out of 26 items, 5 of them by more than 50 percentage points. Six items present slightly decreased performance (decreases ranging from 3.3 to 26.7 percentage points), and 2 items present a bigger decrease (40 and 63.4 percentage points).

*Answering RQ3: How has OpenAI's vision-capable chatbots' performance on the TUG-K evolved since October 2023?*

We found that ChatGPT-4's performance changed on several items between October 2023 and April 2024, with some items improving and some worsening for no clear reason. However, on average, the

TABLE 2 Performance of ChatGPT's different versions on the TUG-K, with the items (column 1) ranked by best to lowest performance according to "ChatGPT-4 (April 2024) No ADA" (grey column).

| Item | ChatGPT 4 (Oct 2023) | ChatGPT 4 (Apr 2024) | | ChatGPT 4o |
|---|---|---|---|---|
| | | No ADA | ADA | |
| 20 | 51.7 | 100 | | 93.3 |
| 10 | 100 | 100 | | 100 |
| 19 | 100 | 93.3 | | 100 |
| 26 | 58.3 | 86.7 | | 100 |
| 22 | 95 | 83.3 | | 100 |
| 16 | 13.3 | 76.7 | 70 | 13.3 |
| 2 | 30 | 63.3 | | 50 |
| 12 | 68.3 | 56.7 | | 40 |
| 6 | 75 | 53.3 | 43.3 | 63.3 |
| 3 | 100 | 50 | | 100 |
| 17 | 5 | 43.3 | | 100 |
| 9 | 10 | 40 | | 0 |
| 25 | 71.7 | 40 | | 100 |
| 4 | 10 | 36.7 | 40 | 10 |
| 14 | 81.7 | 36.7 | | 16.7 |
| 21 | 45 | 33.3 | | 60 |
| 18 | 21.7 | 26.7 | 26.7 | 66.7 |
| 24 | 1.7 | 26.7 | | 100 |
| 11 | 3.3 | 20 | | 33.3 |
| 7 | 58.3 | 13.3 | 3.3 | 43.3 |
| 13 | 1.7 | 10 | | 56.7 |
| 8 | 33.3 | 10 | | 100 |
| 5 | 6.7 | 3.3 | 3.3 | 0 |
| 15 | 5 | 0 | | 20 |
| 1 | 6.7 | 0 | | 23.3 |
| 23 | 31.7 | 0 | | 33.3 |
| Average | 41.7 | 42.4 | 40.9 | 58.6 |

Column 2 reports the data collected in October 2023. Column 3 reports the data collected in April 2024 and it is further divided into columns showing performance with the Advanced Data Analysis plug-in disabled ("No ADA") and enabled ("ADA"). The empty cells indicate that the ADA was not used for those items, so the performance is the same as in "No ADA." The rightmost column reports the data from the newest freely available version of the chatbot, ChatGPT-4o, collected in May 2024.

performance on the test as a whole changed only marginally from October 2023 to April 2024. Furthermore, the use of the ADA plug-in did not have an impact on the overall performance. We can also see that the latest OpenAI chatbot, ChatGPT-4o, outperforms all previous ones. Despite its performance being still far from that of an expert physicist or physics teacher, it stands out from all the other chatbots and makes a big step in that direction.

## 4.4 Other findings

While generating the data for this paper and reading through all the responses from the eight different chatbots, we also recorded some qualitative patterns in their behaviour that can be of interest to educators and other users, as well as researchers on the topic.

### 4.4.1 Citing sources

Gemini and Copilot often refer to external internet sources and sometimes even directly cite them (i.e., Wikipedia, tutorial videos on YouTube, student forums and homework help sites). In particular, Gemini and Copilot often refer to Chegg (a website where users upload exam questions and solutions) and, in some cases, include a link to a scan of the entire TUG-K with marked correct solutions. However, somewhat surprisingly, despite having access to it, both chatbots still performed poorly on the test.

### 4.4.2 Style issues

In a few responses (1%), Gemini 1.0 Pro and Ultra write parts of their responses in a language different from that of the English prompt (e.g., Chinese, Arabic), together with its translation in brackets, or use formulations in different languages (e.g., Japanese, Thai). For example:

> "The best interpretation of the graph in the image is (E) The object is moving with a uniformly increasing velocity.
>
> In a velocity vs. time graph, the slope of the line represents the acceleration of the object. A positive slope indicates that the object is accelerating, which means its velocity is increasing. A constant positive slope, like the one in the graph, indicates that the object is undergoing uniform acceleration.
>
> 需要注意的是，速度不是位移。位移是指物体从一个位置移动到另一个位置的总距离，而速度是物体在给定时间内移动的速率。
>
> (Note that velocity is not displacement. Displacement refers to the total distance an object moves from one location to another, while velocity is the rate at which the object is moving at a given time.)" [Gemini 1.0 Ultra, response 17–02].

Copilot has the most informal style of the tested chatbots and is the only one that regularly uses emojis in its responses. It is also the only chatbot that produces attempts at humour, typically in the last line of its responses, although its jokes mostly felt somewhat out of place, as they were difficult to contextualize with the rest of the response. For example:

> "The graphs that represent motion at constant velocity are I and III. So, the correct answer is B) I and III. Here's a brief explanation of each graph: […] Remember, this is a fun comparison, and bananas are not typically used as a unit of measurement for height! 🍌😊" [Copilot, response 12–06].

In 23.7% of its responses, Claude 3 Sonnet stated that its responses are generated without reproducing copyrighted material, suggesting that it was trained to be cautious about content reproduction. Here is an example:

> "The graph that represents an object's motion with a uniformly increasing velocity is graph III. This can be deduced from the straight line graph with a positive, increasing slope, indicating a constant rate of acceleration over time. To answer the multiple choice question without reproducing any copyrighted material, the correct choice is E) III only." [Claude 3 Sonnet, response 26–04].

We have found that for some chatbots, repeated prompting on an item produces similar answers within the same chatbot, and in a few cases the answers were precisely the same. This mainly happens for Gemini and Copilot, suggesting that their temperature parameter setting may be lower than for other chatbots.

# 5 Discussion

## 5.1 Implications for education

While LLM- and LMM-based chatbots have great potential within the educational domain, they can also hinder learning if used improperly. Given their far-from-perfect performance on basic graphical analysis tasks, we can say that their use presents potential risks to students' development of understanding of subject content, critical thinking and problem-solving skills.

Before we began this study, we expected that freely available chatbots with vision capabilities would perform worse than subscription-based chatbots. If this were the case, then students from underserved communities could experience negative impacts due to the use of such technology in education, both because they are less likely to be able to afford these services themselves and because the schools they attend are less likely to provide subscription-based resources to their students. This phenomenon could present a new contribution to the technological divide that already exists between the privileged and underserved communities (Servon and Nelson, 2001). On the other hand, the availability of state-of-the-art technology can also positively impact underserved communities (Mhlanga, 2023).

This study found that a divide between different chatbots exists, however, it is not related to a performance gap between freely available and subscription-based chatbots. Instead, the performance divide is mainly a split between ChatGPT and other chatbots. In particular, the freely available ChatGPT-4o outperforms even its subscription-based counterpart, ChatGPT-4. This extends the possibilities for using vision ability in the educational context even in populations that cannot afford subscription-based services. However, as technology continues to evolve, there is no guarantee that a technological divide driven by access to subscription-based services will not emerge in the future. Educators and education researchers should remain vigilant and continue exploring the capabilities of different chatbots to spot differences that may cause future divides. However, for the time being, the state-of-the-art chatbot, ChatGPT-4o, remains freely available to all users.

As we have shown, there are large differences between freely available chatbots, with some of them exhibiting very poor overall performance, despite them being advertised for use in tasks involving interpretation of image input. We would, therefore, advise against using these versions of chatbots for educational purposes. However, it is also crucial to acknowledge that even the top-performing chatbot, ChatGPT-4o, still makes mistakes. Interestingly, these mistakes are not typical of humans, and they cannot be accounted for by the two categorisation systems presented in this paper (conceptual and procedural). This also limits its usefulness as a model of a student for the purpose of educational material testing or assessment development. Its unreliability and unpredictability also present a severe limitation for high-stakes applications, such as grading students' work or assisting vision-impaired students.

Despite still not performing at the level of experts, the observed improvements in the performance of OpenAI's chatbots suggest that the technology may soon become useful in some educational contexts that require the interpretation of graphical representations. This development has important implications for AI tutoring systems, which may soon be able to reliably interpret student-drawn representations. Additionally, this technology holds promise for improving accessibility for vision-impaired students.

## 5.2 Limitations and future work

Our findings are based on the chatbots' performance on the TUG-K, which provides only a particular kind of visual representation (graphs) from a very specific domain (kinematics). Therefore, the results may not represent the chatbots' ability to interpret other scientific representations more generally. Exploring other forms of representation may offer insights into what features make an image interpretable by a chatbot. Neither the categorization based on content-based knowledge objectives nor the procedural nature of items revealed overarching performance patterns. Both categories offer a straightforward and natural way to categorize test items from a "human" perspective, suggesting that chatbot abilities should not be directly compared to those of human students. We hypothesize that even more fine-grained visual aspects of the items may be crucial in determining how well chatbots can interpret and respond to them. Future research could involve redesigning the graphs, either by hand or using software, to explore the impact of these changes on chatbot performance. An alternative approach could be looking for correlations and clusters in the performance data and analysing the emerging groupings for common characteristics. In doing so, we may discover that such categories hold little to no intuitive meaning in terms of human perception. Furthermore, there may not be many patterns that generalize across chatbots because of the differences in their underlying architecture and training data.

The TUG-K has a specific format as a multiple-choice test, another important factor limiting the generalizability of the findings. The inclusion of the five options in the screenshots—and thus in the prompt—is likely to influence the chatbot's output. It is not uncommon for a chatbot to pick an option because it is "the closest one that matches the calculations," or "the most correct among the provided ones." This behaviour reflects that of a student choosing an option for the same exact reasons. However, it does not allow the chatbot to rely solely on "its reasoning" to solve the task. The impact can be more or less critical depending on the nature of the options, which could be text, numerical data, or graphs. It may be valuable to explore the chatbots' performance on the TUG-K excluding the answer options, even though this would require reformulating most of the tasks.

Another line of research could look into the impact of different prompting techniques on chatbots' performance. Despite being challenging, several prompt engineering techniques have proven effective for reasoning tasks. In particular, providing the chatbot with adequate context and instructing it about the role it is supposed to play can lead to better results (Polverini and Gregorcic, 2024a). Testing these techniques on tasks that involve visual interpretation is an interesting avenue for future research.

## 6 Conclusion

This study evaluated the performance of eight large multimodal model (LMM)-based chatbots on the Test of Understanding Graphs in Kinematics (TUG-K), to inform their potential application in STEM and medical education.

Our findings challenge the assumption that subscription-based models are inherently superior. Despite being freely available, ChatGPT-4o outperformed not only other chatbot models, but also its subscription-based counterpart, ChatGPT-4. This holds promising implications for AI tutoring systems in topics requiring interpretation of graphical representations and holds great potential for improving accessibility for vision-impaired students. However, the persistence of unpredictable errors even with ChatGPT-4o indicates that reliance on such tools for high-stakes applications like tutoring, grading, or assisting in exams requires caution and may still be premature.

Categorizations of items by content-based knowledge objectives and procedural nature revealed few patterns, indicating that chatbots' abilities should not be directly compared or mapped onto those of human students. This also limits chatbots' utility as a model of a student, e.g., for the purpose of developing assessments.

LMM-based chatbots hold considerable potential to transform educational practices. However, careful consideration and continued research beyond commercial interests are essential to fully realise this potential, avoid pitfalls, and address existing challenges.

## Data availability statement

The datasets presented in this study can be found in online repositories. The names of the repository/repositories and accession number(s) can be found below: https://doi.org/10.5281/zenodo.12180242, https://doi.org/10.5281/zenodo.11183803, and https://doi.org/10.5281/zenodo.12179689.

## Author contributions

GP: Conceptualization, Data curation, Formal analysis, Investigation, Methodology, Project administration, Software, Validation, Visualization, Writing – original draft, Writing – review & editing. BG: Data curation, Formal analysis, Investigation,

Methodology, Project administration, Supervision, Validation, Visualization, Writing – original draft, Writing – review & editing.

# Funding

The author(s) declare that no financial support was received for the research, authorship, and/or publication of this article.

# Conflict of interest

The authors declare that the research was conducted in the absence of any commercial or financial relationships that could be construed as a potential conflict of interest.

# Publisher's note

All claims expressed in this article are solely those of the authors and do not necessarily represent those of their affiliated organizations, or those of the publisher, the editors and the reviewers. Any product that may be evaluated in this article, or claim that may be made by its manufacturer, is not guaranteed or endorsed by the publisher.

# Supplementary material

The Supplementary material for this article can be found online at: https://www.frontiersin.org/articles/10.3389/feduc.2024.1452414/full#supplementary-material